# Towards room-temperature single-layer graphene synthesis by $C_{60}$ Supersonic Molecular Beam Epitaxy

Roberta Tatti°, Lucrezia Aversa°, Roberto Verucchi°*, Emanuele Cavaliere[+], Giovanni Garberoglio[#], Nicola M. Pugno[†‡|], Giorgio Speranza[|], Simone Taioli*[#%].

°CNR, Institute of Materials for Electronics and Magnetism (IMEM), Sede di Trento (Italy)

[+]Dipartimento di Matematica e Fisica "Niccolò Tartaglia", Università Cattolica del Sacro Cuore, Brescia (Italy)

[#]European Centre for Theoretical Studies in Nuclear Physics and Related Areas (ECT*), Bruno Kessler Foundation & Trento Institute for Fundamental Physics and Applications (TIFPA-INFN), Trento (Italy)

[†] Laboratory of Bio-inspired & Graphene Nanomechanics, Department of Civil, Environmental and Mechanical Engineering, University of Trento (Italy)

[‡]School of Engineering and Materials Science, Queen Mary University of London (UK)

[|]Center for Materials and Microsystems, Bruno Kessler Foundation, Trento (Italy)

[%]Faculty of Mathematics and Physics, Charles University in Prague (Czech Republic)

**Corresponding Authors**

* taioli@ectstar.eu, Simone Taioli

* roberto.verucchi@cnr.it, Roberto Verucchi

**ABSTRACT** High-kinetic energy impacts between inorganic surfaces and molecular beams seeded by organics represent a fundamental case study in materials science, most notably when they activate chemical-physical processes leading to nanocrystals growth. Here we demonstrate single-layer graphene synthesis on copper by $C_{60}$ supersonic molecular beam (SuMBE) epitaxy at 645 °C, with the possibility of further reduction. Using a variety of electron spectroscopy and microscopy techniques, and first-principles simulations, we describe the chemical-physical mechanisms activated by SuMBE resulting in graphene growth. In particular, we find a crucial role of high-kinetic energy deposition in enhancing the organic/inorganic interface interaction, to control the cage openings and to improve the growing film quality. These results, while discussed in the specific case of graphene on copper, are potentially extendable to different metallic or semiconductor substrates and where lower processing temperature is desirable.

**INTRODUCTION**

The synthesis of graphene thin films in vacuum conditions can be achieved by several approaches.[1] For example, chemical vapor deposition (CVD) on metallic substrates, notably nickel and copper, leads to single-layer graphene epitaxy by exploiting catalytic efficiency of metals.[2] At variance with standard metal-on-metal heteroepitaxy, graphene growth on metals starts at nucleation centers, such as steps or other defects at the substrate surface, and it occurs only at carbon supersaturation of the surface, a clear fingerprint of a large activation barrier for C attachment.[1] Other features make graphene epitaxy on metallic substrates unique. These include the dependence of the growth dynamics on details of the crystal edges, the equivalence between the binding energy of in-plane carbon–carbon bonds (~ 7.4 eV per carbon atom) and of the graphene edge-metal substrate (~ 7 eV per carbon atom), and the reversibility of the growth dynamics.[1] However, high working temperatures,[3] even in excess of 1000 ºC, are needed in CVD to obtain good quality graphene layers and to initiate the desorption of the hydrogen atoms present in the hydrocarbon precursors. Furthermore, graphene growth by CVD may be critically affected by carbon solubility within the bulk and, finally, by the bond strength between carbon atoms and metal surface. Both these factors depend on process temperature conditions and, typically, CVD single-layer graphene exhibits several defects and polycrystalline structure.[4] Thus, much effort is currently devoted to a better understanding of the growth dynamics on substrate surfaces, to achieve large single-domain dimensions, optimal grain boundary matching and lower processing temperature.[4]

In this work, aiming at overcoming these issues, we demonstrate the possibility of inducing $C_{60}$ cage unzipping by supersonic molecular beam epitaxy (SuMBE) on single-crystal (111) and polycrystalline copper surfaces. SuMBE application to graphene growth will be studied by investigating electronic and structural properties of the synthesized $C_{60}$/Cu thin films and the role of thermal energy in single-layer graphene synthesis by a variety of *in-situ* and *ex-situ* experimental methods, such as electron and Raman spectroscopy and scanning microscopy techniques.

Furthermore, first-principles simulations based on density functional theory (DFT) will be used: i) to simulate the $C_{60}$ impact on Cu(111) surface at several kinetic energies (KE); ii) to show the

**ABBREVIATIONS** AES, Auger Electron Spectroscopy; BE, Binding Energy; BO, Born-Oppenheimer; CL, Core Level; DFT, Density Functional Theory; FFT, Fast Fourier Transform; FWHM, Full Width at Half Maximum; IP, Ionization Potential; KE, Kinetic energy; LA, Longitudinal Acoustic; LEED, Low-Energy Electron Diffraction; MBE, Molecular Beam Epitaxy; NAMD, Non-Adiabatic Molecular Dynamics; RT, Room Temperature; SEM, Scanning Electron Microscopy; STM, Scanning Tunneling Microscopy; SuMBE, Supersonic Molecular Beam Epitaxy; TO, Transversal Optical; UPS, Ultraviolet Photoelectron Spectroscopy; VB, Valence Band; WF, Work Function; XPS, X-ray Photoelectron Spectroscopy.

crucial role of non-adiabatic effects on cage breaking; iii) finally, to follow the long-time dynamics after cage rupture leading eventually to graphene formation.

**RESULTS AND DISCUSSION**

**SuMBE deposition of $C_{60}$ on copper, core and valence band characterization of the films**

SuMBE has been already successfully used to grow nanocrystalline cubic silicon carbide (3C-SiC) at room temperature (RT) on a Si(111)7x7 surface from $C_{60}$ precursor.[5,6] By means of a supersonic expansion of a carrier gas (He or $H_2$), precursors can achieve kinetic energies (KEs) up to tens of eV with freezing of internal degrees of freedom. Most importantly, this technique enables chemical/physical processes on the target surface not achievable by molecular beam epitaxy (MBE) and CVD working at thermal equilibrium.[7]

Due to its relative low cost and highest abundance among fullerene's family, fullerene ($C_{60}$) represents an optimal choice as carbon precursor for graphene growth, provided that its cage is unzipped. Furthermore, being only composed of carbon atoms arranged in a $sp^2$ icosahedral-symmetry network of hexagons and pentagons, $C_{60}$ does not contain chemical elements undesired in film growth; indeed, graphene synthesis from $C_{60}$ has been theoretically proposed.[8] In particular, the possibility of retaining the original faceted structure after cage decomposition is a strong stimulus towards using these carbon allotropes for this scope,[9] despite their chemical and mechanical stability limited its actual adoption in graphene synthesis so far.[10-14] For example, $C_{60}$ thermal decomposition on nickel in the 710-825 °C range resulted in the growth of multiple- and single-layer graphene,[11] while graphene nano-islands at ~500 °C and single-layer at ~920 °C were synthesized from $C_{60}$ on Ru(0001).[10] Furthermore, graphene nanostructures have been further obtained from $C_{60}$ by oxidation,[14] upon increasing temperature and pressure,[13] and by $C_{60}$ cage unzipping via annealing in hydrogen at temperatures above the stability limit.[12] Finally, graphene synthesis has been reported more widely by using high-impact collision of carbon nanotubes on several substrates,[15-16] owing to the higher probability to unzip this carbon allotrope.

In this work, RT $C_{60}$ film growth by SuMBE was performed at 15 and 35 eV KE on both Cu(111) single crystal and Cu poly, with post-deposition annealing at different temperatures. To analyze the results of $C_{60}$ high-impact collision on a copper surface, we performed *in-situ* electron surface spectroscopy measurements (details on the experimental set-up is found in the Material and Methods section). Initially, for comparison with MBE experiments, we deposited a $C_{60}$ 20 nm film at 15 eV KE on Cu poly at RT. The C1s core level (CL) from $C_{60}$ 20 nm film (Figure 1. a(1)) is characterized by a main symmetric component, located at 284.60 eV (FWHM = 0.80 eV, 87% of total C1s area), typical of C-C $sp^2$ bonds and by loss/shake up structures at higher binding energies (BEs).[17] Valence band (VB) in Figure 1.c(2) is dominated by several features, with the highest occupied molecular orbital (HOMO) being located at ~2 eV. We evaluated from UPS analysis a work function (WF) of 4.80 eV, with an ionization potential (IP) of 6.5 eV.

The $C_{60}$ 1 monolayer (1ML) was obtained by annealing the 20 nm $C_{60}$ film at 400 °C on a Cu poly, removing all physisorbed species. C1s CL analysis shows an asymmetric main peak (Figure 1. a (2)) located at 284.14 eV (FWHM 0.92 eV, 87%), a component labelled P1 in Figure 1.(a) at 283.44 eV (FWHM 1.00 eV, ~3%) and the presence of loss structures at higher BEs. Additionally, the same ~-0.45 eV BE shift and band enlargement as for the $C_{60}$ 20 nm film can be found in VB spectrum (Figure 1.c(3)). Copper 3d bands intensity is decreased with respect to clean surface (Figure 1.c(1)), while main features and Fermi edge are still visible. The WF is ~5.0 eV, higher than for Cu surface (WF ~4.8 eV) and thick $C_{60}$ film. These C1s CL features (BE, width and weight) can be found even in $C_{60}$ 1ML films deposited at RT by SuMBE at 15 and 35 eV precursor KEs on substrates of Cu poly/Cu(111) (Figure 6).

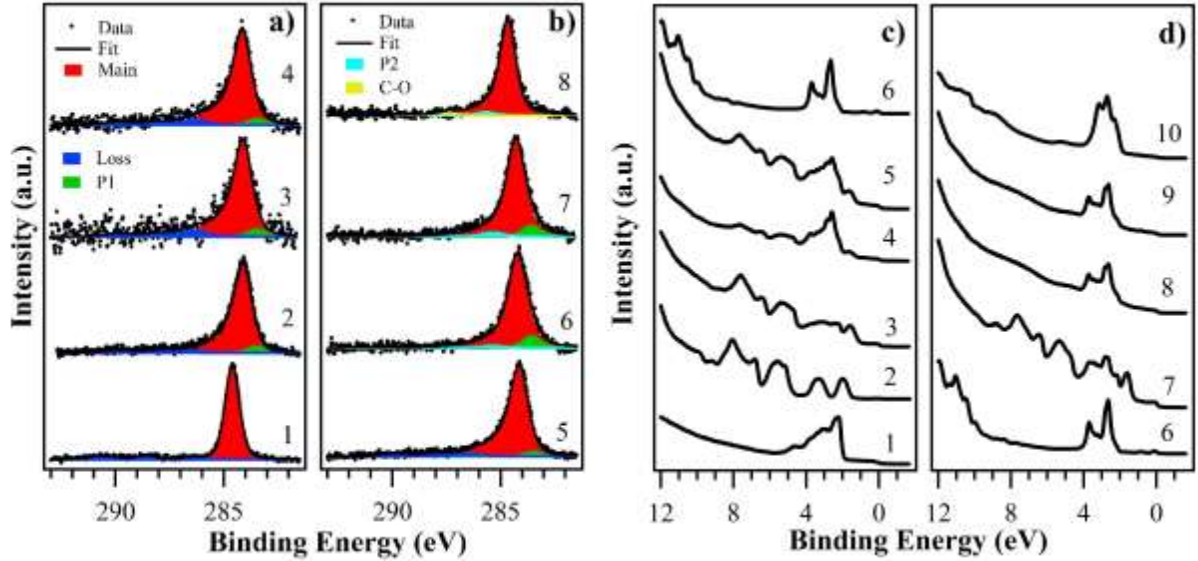

**Figure 1. a)** C1s CL from $C_{60}$ film deposited at RT by SuMBE on Cu poly at KE=15 eV (1, 2) and Cu(111) at KE=35 eV (3, 4) with thickness: 1) 20 nm; 2) 1 ML, after annealing 20 nm film at 400 °C; 3) 0.3 ML; 4) 0.6 ML. **b)** C1s CL from $C_{60}$ 1 ML films deposited at RT for precursor KE=35, after thermal annealing at 425°C (5), 645°C (6), 795°C (7). C1s emission from commercial graphene single-layer on Cu foil is shown for comparison (8). **c)** VB from Cu poly (1); VB analysis of $C_{60}$ films deposited by SuMBE on Cu poly at RT for KE=15 eV (2-3) and Cu(111) at KE=35 eV (4, 5) with thickness: 2) 20 nm; 3) 1 ML, after annealing a 20 nm film at 400 °C; 4) 0.3 ML; 5) 0.6 ML. **d)** VB from Cu(111) (6); VB from $C_{60}$ 1 ML film deposited at RT with KE=35 eV, after thermal annealing at 425°C (7), 645°C (8), 795°C (9). VB from a commercial graphene single-layer on Cu foil (10) is shown for comparison.

VB spectra show differences only related to the copper substrate (Figure S3), and the WF is the same for the $C_{60}$ 1ML on Cu(111), with a small ~0.1 eV increase from the clean surface (WF = 4.94 eV). The observed C1s peak asymmetry and energy shifts are attributed to charge transfer from the metallic substrate to the $C_{60}$ ML in a chemisorption process.[17,18] At variance with MBE deposition, the C1s CL lineshape recorded after SuMBE is not reproducible via a simple Doniac-Sunjic profile and a new peak (P1) has to be introduced to fit properly the data.[17] This feature, while has never been observed in the 1 ML $C_{60}$/Cu(111) system, was found in $C_{60}$ on Ta(110) and related to charge transfer.[19] The -0.7 eV separation in BE between P1 and the main peak is characteristic of covalent bond formation, as for $C_{60}$ on Si.[6] However, this has to be excluded in the case of $C_{60}$ on copper.

The P1 peak, located at BE lower than expected for graphene on copper,[20] could be instead interpreted as the fingerprint of cage rupture, induced in the $C_{60}$ high impact collision on Cu and leading to free-standing graphene flake formation on the surface. Therefore, if this interpretation were correct, the presence of P1 would suggest an energy threshold for cage opening at about 15 eV. Furthermore, the P1 peak intensity did not show appreciable changes in RT deposition of $C_{60}$ on Cu at 15 eV or 35 eV initial beam KE, a surprising indication that $C_{60}$ unzipping is not improved by doubling precursor KE.

To rule out the presence of P1 in the spectrum as a fingerprint of cage rupture, we deposited 0.3 ML and 0.6 ML $C_{60}$ films on Cu(111) at 35 eV KE, as low coverage and high KE represent the conditions for which $C_{60}$ cage rupture would most likely occur. From C1s CL analysis (Figure 1.a(3,4)), P1 represents always ~3-4% of the total C1s area, while VBs differ only for $C_{60}$ features intensity (Figure 1.c(4,5)). Thus, P1 peak can be safely attributed to copper-to-carbon charge transfer and experimental evidence of cage breaking leading to graphene formation upon $C_{60}$ impact at these kinetic energies was not found. Unfortunately, 35 eV is the highest $C_{60}$ KE attainable by SuMBE deposition in our experimental apparatus; thus, one has to rely on *ab-initio* simulations to find such KE threshold for cage rupture and to further investigate the chemical-physical processes occurring during the impact of fullerene with a Cu(111) surface.

**Born-Oppenheimer DFT and non-adiabatic molecular dynamics simulations**

We performed simulations of $C_{60}$ impact on Cu(111) surfaces, with a series of initial kinetic energies in the range 70–210 eV using DFT (Figure 2.a). In a previous work, concerned with SiC growth induced by SuMBE of $C_{60}$ on Si(111), we demonstrated that substrate temperature has very limited effect on cage breaking mechanisms.[21] Thus, we decided to perform DFT simulations at RT. Details on these simulations and the parameters used are given in the SI. The results of these calculations showed no cage breakup for initial kinetic energies of 70 and 100 eV, whereas breakup was obtained at 210 eV. This energy scale is well beyond SuMBE scope.

However, these simulations rely on the validity of the Born-Oppenheimer (BO) approximation, assuming that ionic and electronic motions proceed on decoupled timescales. In $C_{60}$, the 1.6 eV band-gap corresponds to emission in the frequency region of $10^{15}$ Hz, and the collisional time scale in our case – as estimated using BO-DFT calculations – is of the order of few tens of femtoseconds ($\simeq 10^{14}$ Hz). The ratio between nuclear ($\tau_p$) and electronic characteristic times ($\tau_e$) is of the order of $\tau_p/\tau_e \simeq$ 1–10, thus non-adiabatic effects can be expected to be significant in this case. A full treatment of electronic excitations is unfeasible for the $C_{60}$/copper system used in BO-DFT simulations. Therefore, we decided to use a non-adiabatic molecular dynamics (NAMD) approach using $C_{20}$ as impinging molecule by reducing accordingly the initial KE (1/3 of $C_{60}$).[5] The threshold for complete cage breaking is found at 14 eV (corresponding to 42 eV for the equivalent problem of $C_{60}$) confirming our experimental evidence of a $C_{60}$ rupture KE threshold higher than that attainable by SuMBE. However, large distortions of the $C_{60}$ cage and surface penetration can be found already for KE = 30 eV. We report in Figure 2.b the excited states instantly visited by a $C_{20}$ molecule impinging on the Cu(111) surface on which the system dynamics is evolved and forces are calculated. At C–Cu distance below 2.5 Å (around 20 fs) highly excited energy surfaces become progressively populated until the cage breaks, as is evident in the last frame of the trajectory, reported in Figure 2.c. We conclude that, analogously to what has been observed in the impact of fullerenes on silicon, a model including excited electronic states is necessary to describe accurately the KE threshold of carbon cage rupture, as BO ground-state DFT is in error by a factor of 5.

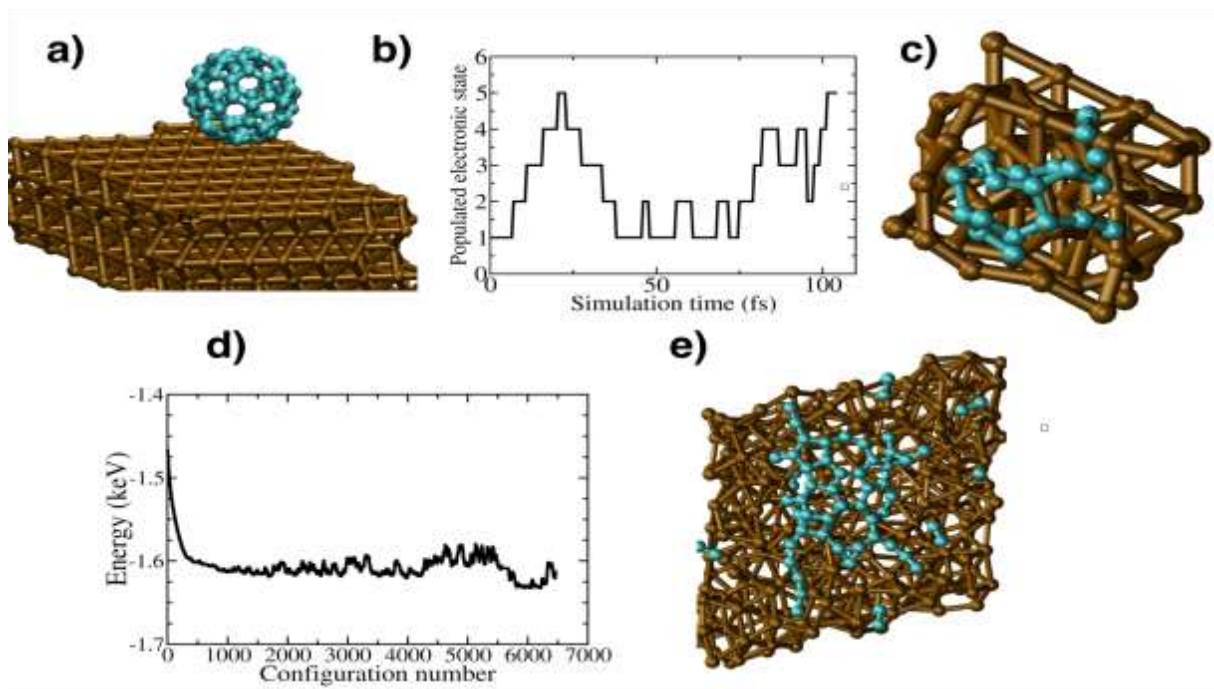

**Figure 2.** a) $C_{60}$ impinging on Cu(111) surface. b) Excited electronic states visited during a TD-DFT simulation of $C_{20}$ impinging on Cu(111) surface at 14 eV (corresponding to $C_{60}$ at 42 eV) c) $C_{20}$ final configuration after cage breaking on Cu(111) surface. d) Total electronic energy of the system during a metadynamics simulation starting from the configuration of a broken $C_{60}$ cage on Cu(111) surface. e) Final configuration of the metadynamics-DFT simulation.

A detailed account of the NAMD simulations performed can be found in the Theory and Calculation section.

**Continuum mechanical model of the high-energy impact**

The kinetic energy threshold for projectile breaking can in principle be estimated by a continuum mechanical model (CM),[22] assuming that this energy is proportional to the object volume $V$. In CM the proportionality constant is the product of the mechanical strength of the projectile and the ratio of the projectile and target densities. The value of the threshold kinetic energy for $C_{60}$ cage breaking is found in fair agreement with our NAMD simulations, assuming a mechanical strength of fullerene close to the value measured in nanotubes. Further details of this approach are reported in the Theory and Calculation section.

### LEED analysis

Experiments and simulations, thus, rule out the possibility of a complete disruption of the cage at the KE achievable by SuMBE of $C_{60}$. One route to follow could be to change the projectile, as a larger mass would result in a higher KE. However, larger mass fullerenes, such as $C_{120}$, are less abundant and more expensive than $C_{60}$ within the fullerene's family. Thus we decided to look for a possible solution by increasing the substrate temperature. In order to evaluate the possible thermally induced $C_{60}$ unzipping we deposited a $C_{60}$ 1 ML film at 35 eV KE on Cu(111) at RT. A thermal annealing sequence in the range 107-795°C (see the Material and Methods section below) has been systematically performed. Furthermore, C1s CL, VB and LEED pattern have been measured, looking for any modifications from $C_{60}$ 1 ML film properties at each temperature.

LEED analysis revealed a diffuse background up to 425°C, when a complex pattern for the 1 ML film appeared (Figure 3.b), with several extra-spots superimposed to the original Cu(111) features (Figure 3.a). The appearance of these signals is explained as a rearrangement of $C_{60}$ molecules on Cu(111) surface with a 4x4 superstructure pattern,[23] made favorable by the very low mismatch (~2%) between the lattice parameter of the organic crystal (10.02 Å) and of the quadruple value of Cu (10.24 Å). This system, characterized by a charge transfer from Cu and $C_{60}$ rotation, undergoes a reconstruction in which a seven atom cavity is formed in the first copper layer, where a single $C_{60}$ cage can be hosted.[24]

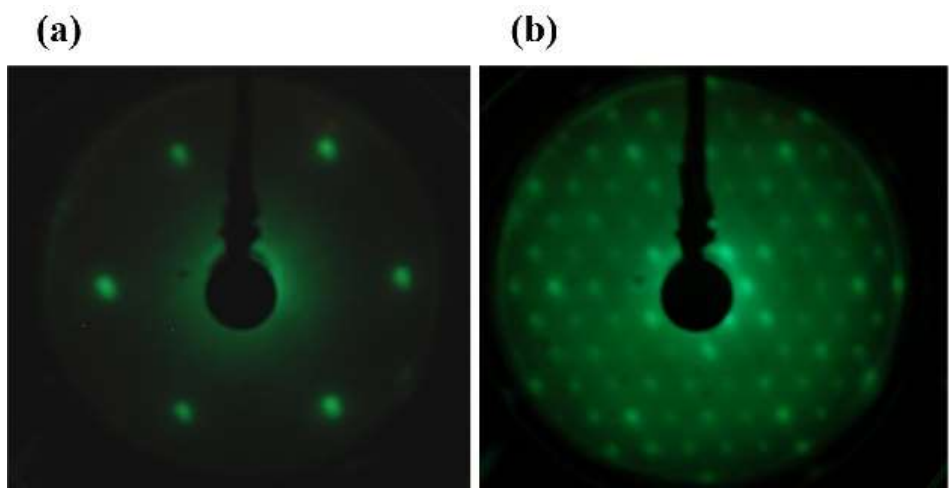

**Figure 3.** LEED pattern (70 eV) for a clean Cu(111) surface (a) and for 1 ML $C_{60}$/Cu(111) (b) after annealing of a 20 nm thick film at 425°C.

**Core-level characteristics of the thermally-assisted SuMBE grown graphene**

A quarter of the entire $C_{60}$ molecule can be accommodated in this cavity and, thus, is found at short distance from 12 copper atoms,[24] leading to the observed chemisorption process and charge transfer from $C_{60}$ to Cu(111) surface. A similar metal surface reconstruction was found in $C_{60}$/Ru(0001) adsorption,[24] where Ru-C strong interaction within the cavity leads to a distortion of $C_{60}$ bonds, to the observed cage opening within the fault line at high temperature and, eventually, to the creation of graphene quantum dots.[10]

In light of both 4x4 relaxation pattern and metallic substrate reconstruction, our interpretation of the C1s P1 peak appearance in CL spectra (Figure 1. a(2-5)), despite showing features of $C_{60}$ 1 ML as found in MBE deposition, is thus to relate it to the carbon atoms experiencing the shortest distance from Cu. While simulations predict indeed cage rupture at KEs out of reach by SuMBE (> 42 eV), however NAMD simulations pointed out that already in the 15 to 35 eV KE regimes RT collisions induce cage distortion and significative surface penetration. We note that this process takes place during RT deposition and before the 4x4 $C_{60}$ rearrangement on the surface, occurring only at temperatures of 100°C higher than those used in MBE deposition (where $C_{60}$ remains intact on copper due to the low KE reached). This means that carbon atoms are in tight contact with the copper surface and much energy has to be spent to diffuse and rearrange the $C_{60}$ cages, partially deformed or in tight contact with copper within the surface cavities.

Chemical properties from C1s CL and VB remain unchanged up to 645°C (Figure 1.b(6)), where C1s CL showed a -10% intensity reduction and deep lineshape change. A comparison with the 1ML CL (Figure 1. b(5)) shows that the main peak is larger (FWHM +0.1 eV), located at higher BE (+0.1 eV) and characterized by a different asymmetry with a typical Doniac-Sunjic lineshape. Furthermore, the previously observed loss structure is absent, the P1 peak is more intense (~8%) and it shows the same energy shift of the main component. A new weak component (P2, Figure 1.b(6)) is present at 285.35 eV (FWHM ~1.25 eV, ~3-4%). These features are typical of a defected graphene single-layer.[25] C1s CL analysis from a commercial graphene single-layer on a copper foil (Figure 1. b (8)) is characterized by a main peak located at 284.65 eV (FWHM 0.88 eV), a peak at 285.65 eV (P2, FWHM 1.00 eV) and a further component at 287.40 eV (FWHM 1.30 eV) due to presence of C-O bonds. Thus, apart from a +0.1 eV shift and lower peak broadening, C1s CL suggests we have synthesized a graphene single-layer, revealing the presence of some defects as evidenced by the larger width and very intense P2 peak.[25] It is worth noting the absence of oxidized species in our film, owing to both SuMBE approach and to the use of $C_{60}$ as precursor. VB curves (Figure 1.d(8)) have lost the typical $C_{60}$ features, showing only the Cu 3d band and two broad features in the 6-9 eV region, which is a clear indication of cage rupture occurrence. No significant changes (for C1s CL and VB) were observed up to 795°C (Figure 1. b(7) and 2.d(9)), a behavior that reinforces our confidence of having obtained a stable graphene sheet. Furthermore, our electron spectroscopy analysis, in agreement with that of standard films grown by CVD, can hardly reveal the presence of graphene. A clear LEED pattern, differently from what we obtained for the 4x4 $C_{60}$ reconstruction, was not found indicating the presence of a material with a small coherent length (less than 20 nm) that hinders the formation of LEED diffraction.

**Metadynamics simulations**

To show that once the fullerene cage is unzipped after thermal treatment we expect the carbon atoms to begin reordering in a graphene-like arrangement, we performed first-principles simulations on timescales much longer than those accessible with ab-initio molecular dynamics. In particular, we tried to investigate whether this was the case by exploiting metadynamics,[26] which is a method developed to accelerate the sampling of the configuration space (see the Theory and Calculation section for details or Ref. 27 for a discussion of graphene growth on copper via a Kinetic Monte Carlo approach). We do not expect excited state dynamics to be significant in the rearrangement of C atoms on the Cu surface. In fact, once the cage is broken, dissipation processes will begin to play a significant role and light electrons will quickly relax to their ground state (for a given position of the nuclei). The subsequent nuclear relaxation, leading to C atoms rearrangement, will therefore be mostly determined by the ground-state electronic surface. To investigate the motion of C atoms, we started from a broken $C_{60}$ configuration on top of a Cu(111) surface, and assumed that the subsequent evolution of the system could be described with BO dynamics. Our findings indicate a pronounced tendency of a broken $C_{60}$ on top of copper to rearrange into a graphene-like network. In metadynamics evolution, the number of C–C bonds increases from ~40 to 60. This trend is accompanied with a pronounced decrease of the electron energy – see Figure 2.d – which indicates indeed the exploration of progressively lower energy states. The very high computational cost of performing metadynamics simulations prevented us to go beyond ~6000 BO steps. Nevertheless, even this time-limited dynamics indicates that a broken $C_{60}$ cage on Cu(111) shows the tendency to rearrange in the direction of producing graphene-like structures with carbon molecules hosted within the defected and terraced Cu(111) surface. In particular, we show in Figure 2.e the last frame of our metadynamics simulation (a full trajectory movie is attached to the SI), where one can see the presence of 3 hexagons and a Stone–Wales defect (made by a pentagon and a heptagon). Comparison with similar calculations starting from $C_{20}$ broken cages, where no tendency to form graphene was observed, indicates that the formation of graphene needs a sufficiently high density of carbon atoms on the surface.

**SEM/STM analysis of the samples**

Scanning electron (SEM) and tunneling microscopies (STM) *ex-situ* analysis revealed presence of terraces (Figure 4.a), typical of Cu(111) surface. High resolution STM images (Figure 4.b) show a graphene lattice, presenting dark point defects and bright contrast lattice distortion, separating few nm extended graphene-like domains as confirmed by FFT analysis of the STM data (Figure 4.e). These results highlight definitely the presence of defected single-layer graphene, with a small coherent length that hinders the formation of LEED diffraction. The graphene flake dimensions and high density once more substantiate our view of thermally-assisted unzipping of $C_{60}$ molecules arranged on the top of Cu(111) surface due to the SuMBE deposition.

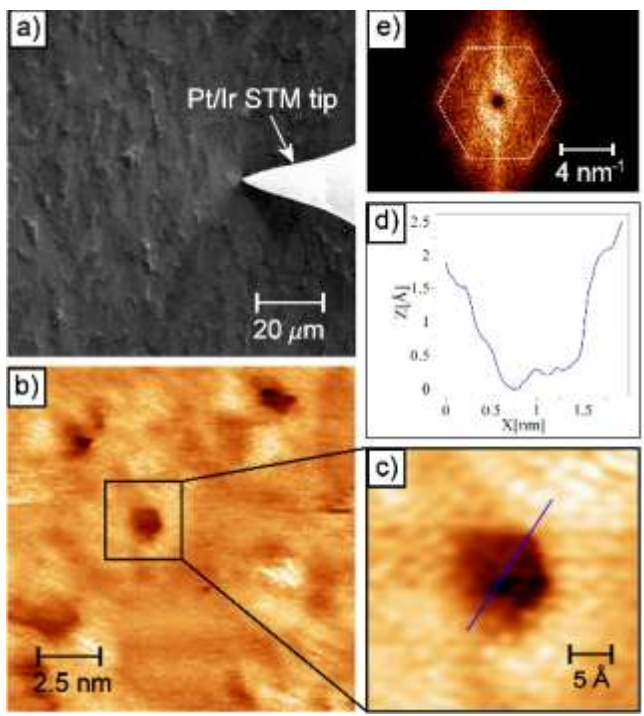

**Figure 4.** $C_{60}$ 1ML on Cu(111) after annealing at 645°C. SEM (a) and STM (b-e) analysis. c) details of a dark region, with line profile (d); e) FFT analysis of the image in b).

Dark region line profile reveals a depth of ~2 Å and a width of ~1 nm (Figure 4. c,d). As Cu(111) interlayer distance is 2.06 Å, dark regions are compatible with the formation of fault lines on the top of the metal representing the $C_{60}$ adsorption sites, which, as mentioned before, are created by a seven atom vacancy. Once unzipped, the bottom part of the cage remains inside the first layer, being responsible for the P1 peak signal in CL spectra of our system, still present after graphene synthesis (Figure 1.b(6, 7)).

**Raman analysis of the samples**

The ultimate evidence for demonstrating graphene synthesis is given by Raman analysis. As shown in Figure 5. (a), spectra from three different zones are characterized substantially by the same features, dominated by the G (E2g mode) and 2D (second order Raman scattering process in curved graphene) bands at 1589 cm$^{-1}$ and 2665 cm$^{-1}$. G band is biphasic and is formed by a second component at 1624 cm$^{-1}$. D band in defected graphene is generally split in two (D and D' bands), located at approximately 1350 cm$^{-1}$ (not shown in our spectra) and 1620 cm$^{-1}$, respectively. A minor peak is found at ~2400 cm$^{-1}$ (Figure 5. (a)). This can be assigned to the G* band,[28] while at ~2470 cm$^{-1}$ appears a weak feature usually associated to the D+D" band.[29] Again, D+D' band is visible at ~2900 cm$^{-1}$.[30] All these features can be associated to the presence of a certain degree of disorder in our single-layer graphene. The G* band can be explained by an inter-valley process involving one TO and one LA phonon (TO and LA are one of the three optical and acoustic phonons present in the phonon dispersion of graphene). The main difference between the acquired spectra is given by a modulation of the 2D/G peak intensity, equal to 3.07, 1.59, 3.87 for zones A, B, C, respectively (Figure 5. (a)).

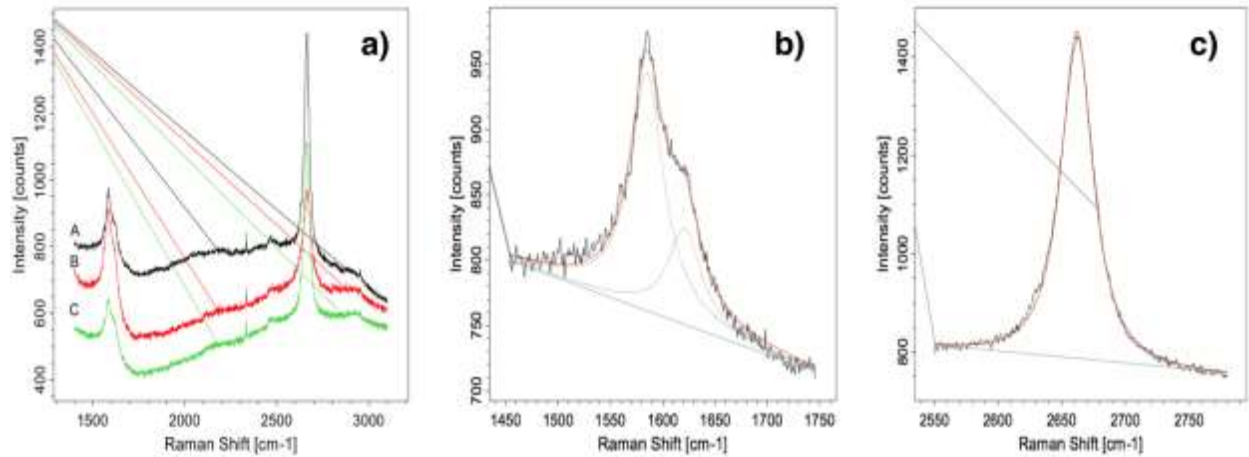

**Figure. 5** Micro Raman analysis of $C_{60}$ 1ML on Cu(111) after annealing at 645°C. a) wide Raman spectra acquired in different regions of the sample. b) an example of high resolution G+D' and (c) 2D bands acquired on the same sample, together with their Lorentzian peak fitting.

The decrease in $I_{2D}/I_G$ ratio reflects energy dispersion through larger number of channels, i.e. a high degree of non-crystalline phases. Indeed, the introduction of disorder through high energy electron beam irradiation of graphene leads to different trends of $I_D/I_G$ and $I_{2D}/I_G$ intensities.[30] In particular, in the low disorder regime, $I_D/I_G$ increases while the opposite trend can be found at high dose of energetic electrons (high disorder regime). Differently, the $I_{2D}/I_G$ ratio assumes a descending trend from the initial irradiation stages revealing insensitivity to the high/low disorder regimes. Finally, it is known that structural information on graphene, in particular the presence of single- or multi-layers, is reflected by 2D features. In our case, the G+D' and 2D features were fitted to follow changes in their morphology by changing the acquisition regions (A), (B) and (C) of the sample surface. An example of fit performed in region (A) is shown in Figure 5 (b,c) (fits of regions (B) and (C) provided in the SI). Only a Lorentzian component is needed to fit the 2D peak (Figure 5. (c)). This information along with the 2D lineshape visible symmetry leads us to the conclusion that we are indeed analyzing a single-layer of graphene.

**CONCLUSIONS**

In conclusion, in this work we report single-layer graphene growth by thermal decomposition of $C_{60}$ films deposited by SuMBE on Cu(111) surfaces. To the best of our knowledge, while MBE thermal decomposition of graphitic layers was already achieved,[31] a single-layer graphene from $C_{60}$ has never been synthesized. SuMBE approach, inducing a tighter $C_{60}$ adsorption within the copper surface, creates favorable conditions for cage unzipping via thermal processes with respect to other widely used approaches. Cage opening, in particular, was not achieved at the KE attainable by SuMBE of $C_{60}$, in agreement with non-adiabatic molecular dynamics simulations, predicting breaking well above 40 eV. However, cage unzipping has been obtained by thermal treatment of $C_{60}$ deposited by SuMBE, after a reconstruction of the surface allowing for an effective molecular orbital/metal valence states overlap. The possibility to control cage rotation, supramolecular organization and unzipping process on Cu surface is promising for the reduction of defects in the graphene layer, through a coherent matching of the different flakes originating from $C_{60}$. Nevertheless, the presence of hexagon/pentagon networks in SuMBE-grown graphene, as shown by our STM and Raman analysis and predicted by long-time metadynamics simulations, could be interesting for microelectronic applications, where a band gap has to be induced in graphene, and to study penta–graphene, a new carbon allotrope recently proposed in theoretical studies.[32] Finally, we envisage that graphene synthesis could be induced at RT also during the molecule/surface high impact collision on copper by introducing some impurities on the surface to avoid that the excess of kinetic energy made available by SuMBE is spent in molecular diffusion on the surface rather than in cage breaking. We devise that such approach can be used to synthesize graphene on substrates different from copper, for example directly on semiconductors at a temperature much lower than graphene growth on SiC, and, due to the collimated nature of the beam in SuMBE, in an unprecedented region-selective modality.

**MATERIAL AND METHODS**

**Experimental growth and analysis apparatus**

Experiments were carried out in an Ultra High Vacuum (UHV) system composed of a SuMBE apparatus and main µ-metal chamber (also referred as "analysis chamber"), where is possible to perform a complete *in-situ* film characterization in a clean and controlled environment with a base pressure ($P_{base}$) of $6\times10^{-11}$ mbar. The SuMBE apparatus is composed of a first chamber that holds the supersonic beam's source, with a base pressure $P_{base}$ of $1\times10^{-7}$ mbar, and a second chamber working as a differential pumping stage to better match the SuMBE and UHV vacuum conditions. The high directionality of supersonic molecular beams allows to link directly the two systems during growth without breaking the vacuum in the main chamber (max pressure of $10^{-7}$ mbar during deposition). The $C_{60}$ source, held in the source chamber, is essentially made of two coaxial quartz capillary tubes with an aperture at the end (the "nozzle" characterized by a diameter of about 50μm) and is resistively heated by a shielded tantalum foil. To form the supersonic beam, the $C_{60}$ vapors are seeded in a gas carrier, He or $H_2$, which, combined with suitable vacuum condition and nozzle diameter, generates an isoentropic expansion outside the nozzle. The molecular flux is selected in a definite zone of the expansion where the particles are characterized by a velocity greater than that of sound (Mach number greater than 1). The resulting fullerene beam is characterized by a kinetic energy that depends on the used buffer gas, its pressure and the source temperature to which the fullerene is evaporated, ranging from 0.1 up to 30-35eV and a growth rate on the substrate of about 0.1 Å/min. The beam energy calibration as a function of seeding buffer gas pressure and temperature has been carried out *ex-situ* in a TOF facility.

Several *in-situ* electron spectroscopies for surface physical/chemical characterization can be performed in the main chamber, such as X-ray Photoelectron Spectroscopy (XPS), Auger Electron Spectroscopy (AES), Low Energy Electron Diffraction (LEED) and Ultraviolet Photoelectron Spectroscopy (UPS). In particular XPS spectra have been taken using Mg kα 1253.6 eV photon energy, while UPS has been performed by means of the HeI photon at about 21.2eV. The electron energy analyzer is a VSW HSA100 hemispherical analyzer with PSP electronic power supply and control, the total energy resolution is 0.80 eV for XPS and about 0.10 eV for UPS. The binding energy (BE) scale of XPS spectra was calibrated by using the Au 4f peak at 84.00 eV as a reference, while UPS binding energies were referred to the Fermi level of the same Au clean substrate. The XPS spectra were also background subtracted using a Shirley background, then plotted against BE. The lineshape analysis was then performed using Voigt profiles. Typical uncertainty for the peak energy positioning amounts to ± 0.05 eV, while the full width at half maximum (FWHM) and the area evaluation uncertainties are less than ± 5% and ± 2.5%, respectively.

**Deposition parameters**

Cu(111) single crystal and polycrystalline substrates were cleaned by cycles of sputtering/annealing, by using an $Ar^+$ ion beam at 0.5 keV and annealing controlled by both a thermocouple clamped near the crystal surface, as well as an external pyrometer. The maximum annealing temperature was equal to the final temperature used in each experiment, in order to avoid presence of any sulphur contamination during the thermal process. The copper surface was considered clean when no presence of contaminants (oxygen or carbon) was revealed by AES and XPS techniques, and for the Cu(111) when a clear LEED pattern was observed. All $C_{60}$ films have been deposited at room or higher substrate's temperature, seeding $C_{60}$ in He or $H_2$. After deposition, in some experiments specific thermal annealing treatments have been performed. Fullerene supersonic beam was directed normal to the copper crystal surface. Attention was pointed to calibrate $C_{60}$ deposition in order to grow a reliable and reproducible 1 monolayer (1ML) thin film. Film growth has been achieved after thermal desorption of a thick film of fullerene, in order to remove the physisorbed molecules and leave only the first interacting $C_{60}$ monolayer. This procedure has been performed on both (111) and polycrystalline Cu surfaces. The SuMBE source parameters for the He or $H_2$ transport gas supersonic beams are ~500°C as working temperature, 1200 mbar as gas transport pressure, leading to final kinetic energies (KEs) of 15 and 35 eV, respectively.

**In situ surface characterization**

$C_{60}$ thin films have been deposited by SuMBE on Cu(111) surface at substrate temperature of 20°C (room temperature, RT), using different precursor KEs. The film coverage has been evaluated combining both AES and XPS results. The Cu2p core level (CL) signal has an attenuation length of about 2 nm in our experimental conditions, so at these low $C_{60}$ coverages the photoemission signal is dominated by contributions coming from the substrate bulk and does not provide useful information. A $C_{60}$ 1 ML film at 35 eV KE on Cu(111) has been deposited at RT. A further sequence of thermal annealing has been performed at 107°C, 165°C, 255°C, 326°C, 380°C, 425°C, 498°C, 547°C, 598°C, 645°C, 695°C, 745°C, 795°C. C1s, VB and LEED analysis have been performed to check any changes at each temperature.

**XPS and UPS analysis**

Figure 6 (left panel) shows the Cu2p CL from polycrystalline and (111) copper surfaces. Emission from both surfaces shows the same characteristics, with presence of a 1/2 - 3/2 doublet located at

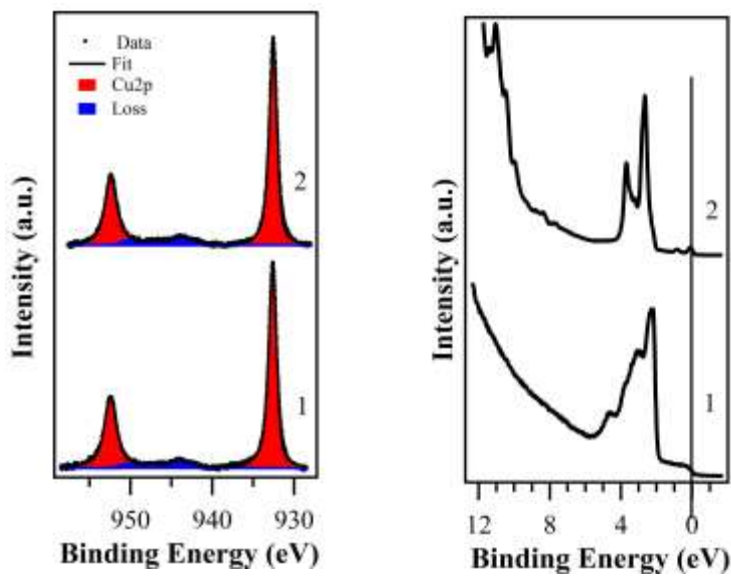

**Figure 6** Cu 2p CL (left panel) and VB (right panel) from a clean polycrystalline copper (1) and from a Cu(111) surface (2).

|          | Cu Polycrystalline |           |      | Cu(111)  |           |      |
|----------|---------|----------|------|---------|----------|------|
|          | BE [eV] | FWHM [eV] | %    | BE [eV] | FWHM [eV] | %    |
| Cu2p 3/2 | 932.60  | 1.14     | 60.7 | 932.54  | 1.13     | 60.9 |
| Cu2p 1/2 | 952.45  | 1.73     | 31.9 | 952.39  | 1.69     | 30.6 |
| Loss     | 942.70  | 2.00     | 1.6  | 942.64  | 2.00     | 1.9  |
| Loss     | 944.30  | 2.00     | 2.7  | 944.23  | 2.00     | 2.9  |
| Loss     | 947.08  | 2.00     | 1.4  | 947.00  | 2.00     | 1.6  |
| Loss     | 950.51  | 2.00     | 1.7  | 950.43  | 2.00     | 2.1  |

**Table 1** Analysis of Cu2p CL from a polycrystalline and a (111) copper surface.

~952.4 eV and ~932.6 eV (see Table 1), with a 19.8 eV BE distance and the expected intensity ratio (1/2). The estimated Cu2p photoelectron attenuation length is about 1.4 nm, thus the low surface sensitivity does not enable an efficient analysis of the last copper atomic layer (0.2 nm), to have evidence of the proposed seven atom vacancy reconstruction.

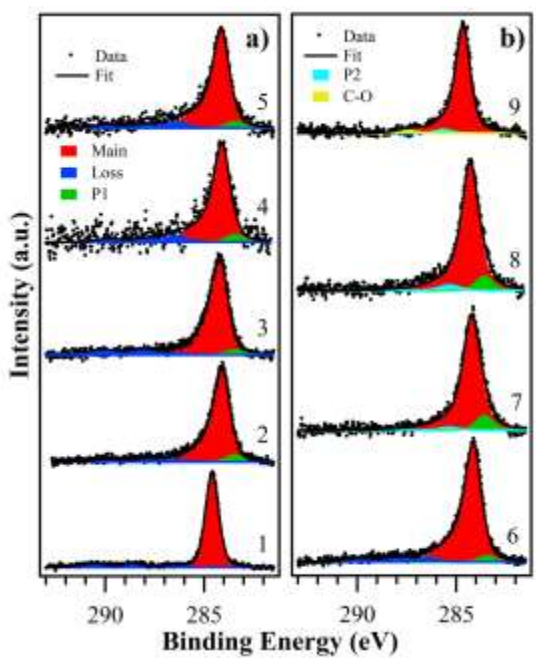

**Figure 7** a) C1s CL from $C_{60}$ films deposited by SuMBE on Cu poly at 15 eV KE (1-3) and Cu(111) at 35 eV KE (4, 5) with thickness: 1) 20 nm; 2) 1 ML, after annealing at 430°C of a 20 nm film; 3) 1 ML; 4) 0.3 ML; 5) 0.6 ML. b) C1s from a C60 1 ML film deposited at RT and precursor 35 eV KE, after thermal annealing at 425°C (6), 645°C (7), 795°C (8). All depositions with substrate at RT. C1s emission from graphene single layer is shown for comparison (9).

On the contrary, valence band (VB) curves are very different for the two copper surfaces, reflecting the ordered structure in the Cu(111) case (Figure 6, right panel). This is evident for the 3d bands, as well as for the Fermi edge region, where presence of surface states dominates VB for the crystalline surface.[33,34] Figure 7 shows all C1s CL from analyzed $C_{60}$ film, while in Table 2 are described component characteristics. As can be seen, C1s core level from all $C_{60}$ 1 ML films are characterized by the same features (within typical errors), apart from film treated at two higher temperatures. Corresponding VB curves are shown in Figure 8.

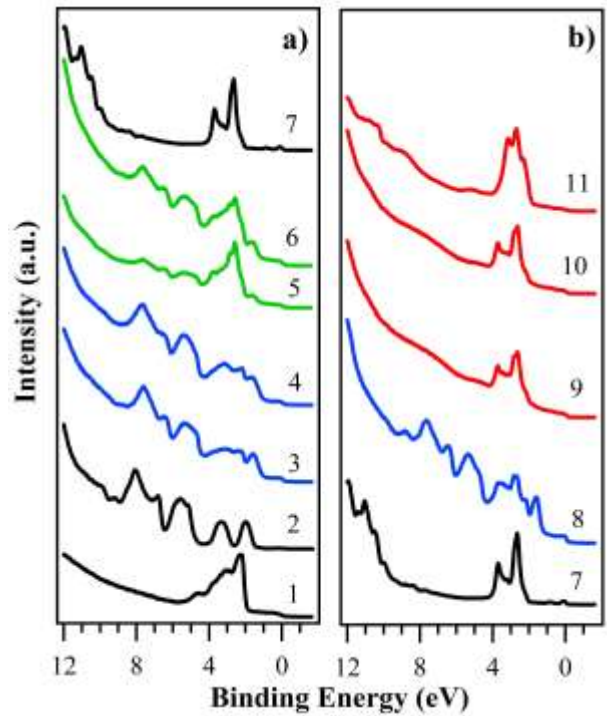

**Figure 8** a) Valence band analysis of $C_{60}$ films deposited by SuMBE on Cu poly at 15 eV KE (2-4) and Cu(111) at 35 eV KE (5, 6) with thickness: 1) 20 nm; 3) 1 ML, after annealing at 430°C of a 20 nm film; 4) 1 ML; 5) 0.3 ML; 6) 0.6 ML. b) C1s from a $C_{60}$ 1 ML film deposited at RT and precursor 35 eV KE, after thermal annealing at 425°C (8), 645°C (9), 795°C (10). All depositions were performed keeping the substrate at RT. VB from Cu poly (1), Cu(111) (7) and graphene single layer (10) are shown for comparison.

|  | C$_{60}$ Bulk | | | 1ML, Poly Cu, KE=15eV, 400°C | | | 1ML, Cu(111), KE=35eV, RT | | |
|---|---|---|---|---|---|---|---|---|---|
|  | BE [eV] | FWHM [eV] | % | BE [eV] | FWHM [eV] | % | BE [eV] | FWHM [eV] | % |
| C-C | 284.60 | 0.80 | 86.6 | 284.12 | 0.90 | 86.9 | 284.14 | 0.92 | 87.6 |
| P1 | 286.50 | 1.00 | 2.0 | 283.44 | 1.00 | 4.5 | 283.44 | 1.00 | 3.1 |
| Loss | 288.60 | 1.30 | 4.6 | 286.41 | 1.30 | 2.1 | 286.41 | 1.30 | 2.8 |
| Loss | 290.60 | 1.50 | 6.8 | 287.95 | 1.50 | 3.5 | 287.97 | 1.50 | 1.6 |
| Loss | 284.60 | 0.80 | 86.6 | 289.90 | 1.50 | 3.0 | 289.92 | 1.50 | 2.9 |

|  | TT @ 425°C | | |  | TT @ 645°C | | | TT @ 795°C | | | Graphene STD | | |
|---|---|---|---|---|---|---|---|---|---|---|---|---|---|
|  | BE [eV] | FWHM [eV] | % |  | BE [eV] | FWHM [eV] | % | BE [eV] | FWHM [eV] | % | BE [eV] | FWHM [eV] | % |
| C-C | 284.14 | 0.92 | 87.6 | C-C | 284.23 | 1.03 | 88.8 | 284.30 | 1.00 | 88.1 | 284.66 | 0.88 | 95.1 |
| P1 | 283.44 | 1.00 | 3.1 | P1 | 283.56 | 1.00 | 8.9 | 283.57 | 1.00 | 7.9 |  |  |  |
| Loss | 286.41 | 1.30 | 2.8 | P2 | 285.35 | 1.22 | 2.3 | 285.35 | 1.30 | 4.0 | 285.66 | 1.00 | 2.7 |
| Loss | 287.97 | 1.50 | 1.6 | C-O |  |  |  |  |  |  | 287.40 | 1.30 | 2.2 |
| Loss | 289.92 | 1.50 | 2.9 |  |  |  |  |  |  |  |  |  |  |

|  | 0.3ML, KE=35eV Cu(111) | | | 0.6ML, KE=35eV Cu(111) | | |
|---|---|---|---|---|---|---|
|  | BE [eV] | FWHM [eV] | % | BE [eV] | FWHM [eV] | % |
| C-C | 284.11 | 0.92 | 84.1 | 284.11 | 0.92 | 85.1 |
| P1 | 283.44 | 1.00 | 4.5 | 283.44 | 1.00 | 3.9 |
| Loss | 286.42 | 1.30 | 5.1 | 286.41 | 1.30 | 4.6 |
| Loss | 287.94 | 1.50 | 3.4 | 287.97 | 1.50 | 3.4 |
| Loss | 289.89 | 1.50 | 2.9 | 289.92 | 1.50 | 3.0 |

**Table 2** Analysis of C1s CL from C$_{60}$ films with different thickness and from 1 ML C$_{60}$ film on Cu(111), deposited at RT by SuMBE at 35 eV after different thermal treatments.

### Ex situ surface characterization (SEM, STM, Raman)

G/Cu(111) samples were investigated by means of a Multiscan Lab by Omicron including the electron column (FEI) for SEM imaging and a room temperature STM. SEM images were collected by SE (secondary electrons) imaging with the FEI electron optics set at 10 keV beam energy and 200pA beam current. The STM images were collected at room temperature with a Pt/Ir tip, prepared by ac electrochemical etching in saturated CaCl2 deionized water solution.

STM was attempted on both the as-grown samples but without success due to high instability induced on the STM tip (Pt/Ir tip) by contamination due to exposure of the sample to the air. Therefore the G/Cu(111) sample was annealed in UHV by radiative heating up to 480 °C, as no significant alteration of this system was expected upon UHV annealing. SEM images were collected before (fig. 1a) and after the annealing procedure (fig. 1b), confirming that no relevant morphological modification of the film occurred after annealing process. STM tip was then positioned with the aid of the SEM during initial tip approach on different graphene flakes.

Raman spectra were acquired with a MicroRaman Aramis (Horiba Jobin-Yvone France) using a 632.8 nm laser wavelength and an air-cooled CCD 1024x256 VIS. The grating used for light dispersion in wide spectra was characterized by 1200lines/mm while 1800 lines/mm was utilized for high resolution of G and 2D band acquisitions. The instrument is equipped with 10x, 50x, 100x objectives. In our experiments spectra were acquired with a 50x magnification.

**Theory/Calculation**

**Born–Oppenheimer Density Functional Theory (BO-DFT) calculations**

The Cu(111) surface was modeled by means of a slab containing 5 Cu layers. The unit cell used in the calculations exposes a Cu surface of 360 Å$^2$ and the length in the orthogonal direction (corresponding to the impact direction) is 25 Å, resulting in a total of 315 Cu atoms.

BO-DFT calculations have been performed using the ab-initio total energy and molecular dynamics program VASP [35-38]. The ion-electron interaction is described using the projector augmented wave (PAW) technique[39] with single particle orbitals expanded in plane waves with a cutoff of 400 eV, which ensures convergence of the electronic structure and of the total energy within chemical accuracy (0.01 eV). Only the Γ point has been used to sample the Brillouin zone, due to the large size of the unit cell.

We tested different exchange-correlation functionals, based on the local density approximation (LDA)[40] or on the gradient-correction expansion (GGA-PBE)[41], finding no effect on the dynamics of the system. All the simulations were then performed using the LDA functional. The temperature adopted during the calculations was 300K, using a Fermi smearing for the electronic population with the same temperature.

Molecular dynamics simulations were performed in the microcanonical ensemble, using a time step of 1fs and integrating the equations of motion for a total of 700 steps. The simulations used an efficient charge-density extrapolation, which speeds up the simulations by approximately a factor of two. The initial condition in DFT simulations was made by juxtaposition of an optimized Cu slab and an optimized $C_{60}$ molecule placed above the surface with a minimum C–Cu distance of 5 Å. A movie of the full trajectory with kinetic energy of 210 eV is available as Supplementary Material.

**Non-adiabatic DFT calculations**

In this approach, the non-adiabatic dynamics is approximated by performing stochastic hops between adiabatic surfaces constructed with the excited states of the system[42], which have been calculated by using time-dependent density functional theory (TDDFT) in the Tamm–Dancoff[43] approximation. Norm-conserving Troullier–Martins pseudopotentials[44] with 11 valence electrons for copper and 4 for carbon were adopted. The electron density is expanded in plane waves, up to a cutoff energy of 1100 eV. The LDA exchange-correlation functional with the Ceperley–Alder[45] parametrization for the correlation has been used.

The nuclei were propagated using Newton's equation of motion on the current adiabatic electronic state and the probability of surface hopping was evaluated by means of the Landau–Zener[46,47] theory. Forces used in the MD simulation are calculated on the adiabatic surfaces populated at the present MD step and constructed with the excited states of the system. These simulations were carried out using the CPMD code[48,49]. Unfortunately, we estimated that the computational cost for simulating the excited-state dynamics of $C_{60}$ impinging on the Cu(111) surface would have been too high to obtain a result in a reasonable time.

Therefore, we considered a smaller yet realistic system. We used a three-layer Cu slab composed of 48 copper atoms, blocking the last layer, with a orthorhombic unit cell exposing a Cu surface of 91.2 Å$^2$ and having a length of 18 Å in the orthogonal direction. Due to the small transverse size of the slab, we chose to simulate the impact of a $C_{20}$ molecule, in order to avoid unphysical interactions with periodic images. The $C_{20}$ was placed initially above the surface so that the closest

C–Cu distance was 4.0 Å. The five lowest lying singlets were included in the calculation of the adiabatic surfaces. All of them were found to be visited during the dynamical evolution of the system. We performed three simulations of $C_{20}$, with initial kinetic energies of 14, 8, and 5 eV (corresponding to 42, 32, and 15 eV for $C_{60}$ having the same initial velocity) to find the kinetic energy threshold. Each simulation lasted 0.2 ps with a time-step of 0.5 fs. The fragmentation of the cage happened only in the first case, while it was not obtained in the other two. As a check, we also performed a BO-DFT simulation of $C_{20}$ impact at an initial kinetic energy of 14 eV, and we did not observe fragmentation, enforcing once more the role of electronic excited states on the cage breaking.

A movie of the fullerene trajectory all the way from the initial condition to breaking is reported in in the Supplementary Material. There, one can clearly observe fragmentation of $C_{20}$ impinging with a kinetic energy of 14 eV on the Cu(111) surface on a timescale spanning 104 fs.

**Metadynamics**

Metadynamics[50] evolves the system according to the usual Newton equations for the nuclei, but adds a history-dependent potential that progressively prevents the system to pass through already visited configurations. In this way, the hopping between metastable basins is faster, exploring the configurations of carbon atoms with an efficiency higher than a usual MD simulation. In our case, the history-dependent potential was made by a series of repulsive Gaussians depending on the coordination-number collective-variable implemented in VASP, with height of 0.5 eV and unit width. This collective coordinate is proportional to the number of C–C bonds in the system. In this way, the simulation is biased towards the breaking of existing bonds and the formation of new ones. A full movie of the metadynamics trajectory is provided as Supplementary Material.

**Continuum mechanical model of $C_{60}$ cage breaking**

The kinetic energy threshold for projectile breaking in a continuum model is proportional to the object volume $V$, where the proportionality constant is the product of the mechanical strength of the projectile and the ratio of the projectile and target densities[52]. The threshold velocity $v$ for breakup at temperature $T$ would then be given by

$$\frac{1}{2}Mv^2 + \frac{1}{2}k_B TNn = \frac{\sigma_f \rho_n}{\rho} V \qquad 1)$$

where $M$ is the mass of $C_{60}$, $\sigma_f$ is the mechanical strength of the fullerene, $\rho_n$ is its density, $\rho = 8960$ kg/m$^3$ is the copper density, $N = 60$ is the number of atoms, $n = 3$ are the internal degrees of freedom per atom and $k_B$ is the Boltzmann constant. In this continuum mechanical model of the impact, the initial velocity could be in principle replaced by a temperature enhancement. However, since the SuMBE deposition effectively freezes the rotational and vibrational degrees of freedom of the beam's molecules while increasing the kinetic energy, the second term in the left hand side of Eq. 1 can be safely neglected.

Assuming $\sigma_f$ of the order of the mechanical strength of carbon nanotubes[51] (~ 50 GPa), the threshold kinetic energy of $C_{60}$ fullerene breakup would be estimated as ~40 eV, in good agreement with the numerical findings.

Notice that an estimation of breakup threshold energy based on the average C–C dissociation bond energy (~4 eV per bond[52]) would result in an expected kinetic energy of at most ~360 eV (corresponding to a complete dissociation of the $C_{60}$ fullerene), which is an order of magnitude higher than the estimate based on the tensile strength. This is however the order of magnitude of cage breakup obtained by the BO-DFT approach (~210 eV).

**Supporting Information**. Movies of the $C_{60}$ trajectory from initial condition to cage breaking are provided in the Supporting Materials.

**ACKNOWLEDGMENTS**

We acknowledge funding from the EU under the FP7$^{th}$ grant agreement 604391 (Graphene Flagship), from FBK via the CMM Director grant "SuperCar", and PRIN project DESCARTES (no. 2010BNZ3F2), MIUR, Italy. N.M.P. is supported by ERC StG Ideas 2011 BIHSNAM (no. 279985), ERC PoC 2013-1 REPLICA2 (no. 619448), ERC PoC 2013-2 KNOTOUGH (no. 632277), and by the Provincia Autonoma di Trento ('Graphene nanocomposites', no. S116/2012-242637 and reg. delib. no. 2266). G.G. and S.T. acknowledge support by INFN through the "Supercalcolo" agreement with FBK. S.T. gratefully acknowledges the Institute for Advanced Studies in Bologna for supporting his ISA fellowship. We thank Prof. D. Alfè (UCL), Dr. S. a Beccara (ECT*), and Dr. M. Dapor (ECT*) for invaluable discussions on simulation topics. Computer simulations were performed on the Archer Supercomputing Facilities (UK) and on the KORE computing cluster at FBK. R.V. thanks Prof. L. Gavioli (Università Cattolica) for his support in SEM/STM analysis, and C. Corradi and M. Pola for their technical assistance.

**Author Contributions**

The manuscript was written through equal contributions of all authors. S. Taioli had the idea and supervised the project with regard to scientific and administrative duties. S. Taioli and G. Garberoglio performed the *ab-initio* simulations. R. Tatti, L. Aversa and R. Verucchi designed, performed the deposition experiments by SuMBE, the *in-situ* characterization of all the samples and analyzed the experimental data. G. Speranza performed the *ex-situ* Raman and electron spectroscopies and analyzed the data. E. Cavaliere delivered the SEM and STM measurements. N. M. Pugno devised the continuum mechanical model of cage breaking. All authors gave approval to the manuscript final version.

**Competing financial interests**

The authors declare no competing financial interests.

**REFERENCES**

(1)   H. Tetlow, J. Posthuma de Boer, I.J. Ford, D.D. Vvedensky, J. Coraux, L. Kantorovich, Phys. Rep. 2014, 542, 195-295